\documentclass[aps,twocolumn,groupedaddress,showpacs,floatfix,amsmath]{revtex4}
\usepackage{tabularx}
\usepackage{graphicx}
\usepackage{dcolumn}
\usepackage{bm}
\usepackage{color}

\definecolor{Remarks}{rgb}{1,0.3,0.3}

\newcommand{\be}{\begin{equation}}
\newcommand{\ee}{\end{equation}}
\newcommand{\bea}{\begin{eqnarray}}
\newcommand{\eea}{\end{eqnarray}}
\newcommand{\etal}{{\it{et~al.\,}}}

\begin{document}

\title{Optimized norm-conserving Hartree-Fock pseudopotentials for plane-wave calculations}

\author{W.~A.~Al-Saidi}
\email{al-saidi@cornell.edu}
\altaffiliation[Present address: ]{Physics Department, Clark Hall,
  Cornell University, Ithaca NY 14850}
\author{E.~J.~Walter}
\affiliation{Department of Physics, College of William and Mary,
Williamsburg, VA 23187-8795}
\author{A. M. Rappe} \affiliation{The Makineni Theoretical Laboratories,
Department of Chemistry, 
University of Pennsylvania, Philadelphia, PA 19104--6323}

\date{\today}

\begin{abstract}

We report Hartree-Fock (HF) based pseudopotentials suitable for
plane-wave calculations. Unlike typical effective core
potentials, the present pseudopotentials are finite at the origin and
exhibit rapid convergence in a plane-wave basis; the optimized
pseudopotential method [A. M. Rappe \etal, Phys.\ Rev.\ B {\bf 41}
1227--30 (1990)] improves plane-wave convergence.  Norm-conserving HF
pseudopotentials are found to develop long-range non-Coulombic
behavior which does not decay faster than $1/r$, and is non-local. This behavior, which
stems from the nonlocality of the exchange potential, is remedied
using a recently developed self-consistent procedure [J.~R.~Trail and
R.~J.~Needs, J. Chem. Phys. {\bf 122}, 014112 (2005)].  The resulting
pseudopotentials slightly violate the norm conservation of the core
charge.  We calculated several atomic properties using these
pseudopotentials, and the results are in good agreement with
all-electron HF values. The dissociation energies, equilibrium
bond lengths, and frequency of vibrations of several dimers obtained
with these HF pseudopotentials and plane waves are also in good
agreement with all-electron results.

\end{abstract}

\pacs{02.70.Ss, 71.15.-m, 31.25.-v}
\maketitle

\section{Introduction}

In nearly all plane-wave density functional
calculations \cite{kohn_nobel}, the use of pseudopotentials is
essential in order to eliminate the atomic core states and the strong
potentials that bind them. This results in replacing the
electron-nuclear Coulomb interaction by
a weaker (often angular-momentum dependent) potential. The advantage is the
reduced computational cost, due to reduction in the number of
electrons and in the plane-wave basis cutoff \cite{pickett}.

In the physics community, pseudopotentials are typically constructed
using density functional theory (DFT) due to its
success in electronic structure calculations \cite{BHS,RRKJ,TM}.
Pseudopotentials based on methods other than DFT are less widely
available. One of the key ingredients of most of these
pseudopotentials is their ``softness,'' meaning that they provide rapid convergence in
a plane-wave basis.  On the other hand, in the chemistry community,
Hartree-Fock (HF) pseudopotentials or effective core potentials (ECPs)
are mostly used \cite{dolg_rev,SBKJC,hay_wadt_psp,
  CLP_psp,dolg_psp}. Most of the available ECPs,
which are typically expressed in a Gaussian basis, are of
limited use in plane-wave calculations, mainly because of their
singular behavior at the origin and slow convergence in a plane-wave basis.

Recently, there has been a growing interest in developing {\emph{soft-core}} HF
pseudopotentials
\cite{greeff_lester,lester_psp,trail_needs1,trail_needs2}.  HF
pseudopotentials have been applied in diffusion Monte Carlo calculations,
and the results suggest that they are better suited for
correlated calculations than DFT-based pseudopotentials
\cite{greeff_lester}. In addition, HF pseudopotentials could be useful
in certain calculations where a negatively charged reference state is needed;
HF tends to bind electrons more strongly  than DFT. Also, hybrid
exchange-correlation functionals, which include some proportion of HF
exact exchange (notably B3LYP~\cite{B3LYP}) have proved successful in electronic
structure calculations, and whether HF or B3LYP based pseudopotentials
would be useful in these type of calculations remains to be explored.

Trail and Needs \cite{trail_needs1} have recently found that
norm-conserving HF pseudopotentials constructed using standard
norm-conserving pseudopotential methods develop a non-decaying and
non-local tail. These pseudopotentials would generally lead to
erroneous results, and even infinite energies in solid state
calculations. Previously, pseudopotentials based on exact exchange
within the optimized potential method (OPM) were also found to develop
a similar, but local and eventually decaying long range structure
\cite{bk_exact_xc,stadele_exact_xc,engel_exact_xc}, which led to
erroneous results in applications.

Different schemes have been proposed to cure the long-range behavior
in the HF and OPM type pseudopotentials mentioned above. Originally
with the OPM method, different groups 
developed procedures for reducing or eliminating the unphysical
long-range behavior, while retaining desired eigenvalue matching
and norm conservation of the pseudopotential
\cite{bk_exact_xc,stadele_exact_xc}. This procedure was later
implemented in a self-consistent manner by altering the
pseudo-orbitals over all space, which led to a small violation of the
norm conservation of the core charge. Nevertheless, the pseudopotentials proved
transferable in both atomic and solid-state calculations
\cite{engel_exact_xc}. Trail and Needs fixed the long-range unphysical
behavior using a similar self-consistent approach \cite{trail_needs1}
that also slightly violates norm
conservation. Ovcharenko \etal \cite{lester_psp} have
recently developed HF-based pseudopotentials which are finite at the
origin and are expressed in terms of Gaussian basis, but are still
relatively hard for typical plane-wave calculations, as we show in
Sec.\ IV.

In this paper, we construct soft-core HF based pseudopotentials
following the Rappe-Rabe-Kaxiras-Joannopoulos (RRKJ) method
\cite{RRKJ}. We use a self-consistent approach to restore the
Coulombic tail in a manner similar to that used by Trail and Needs
\cite{trail_needs1,casino_psp}. Pseudopotentials are highly non-unique entities,
and it is useful to have several alternative forms available with
different properties. For example, one of the advantages of the RRKJ
construction scheme is the softness of the resulting pseudopotentials.

The rest of the paper is organized as follows. We first review
the general construction scheme of the  RRKJ
pseudopotentials, and the procedure we used to ``localize'' the
pseudopotentials (remove the long tail behavior). The crucial steps
in the validation of the pseudopotentials are the study of their
convergence and transferability, {\emph {i.~e.}} how well the pseudopotentials 
converge in a plane-wave basis set, and how well they perform in
environments different than the reference configuration. These
steps are explored in Sections III and IV. In Sec.\ III, we 
study plane-wave convergence of the atoms, and we
investigate the transferability by looking at the ionization energy,
electron affinity, and excitation energy of the first- and second-row
elements. In Sec.\ IV, we study several dimers at the HF level,  and
compare both the all-electron and pseudopotential results. Finally, in
Sec.\ V, we conclude by summarizing the  main points.

\section{Pseudopotential construction} 

We follow the usual density functional theory approach for the construction of
the norm-conserving pseudopotentials starting from an all-electron (AE)
atomic calculation \cite{pickett}. However, in this case, we solve the
Hartree-Fock instead of the Kohn-Sham equations. Our HF
solver is adapted from the code of Ref.~\onlinecite{f_fisher}.

The pseudopotential construction starts by choosing an electronic
reference state, and then solving the Schr\"odinger equation of the
system for the eigenstates $\Phi_{nl}({\bf r})$ and eigenvalues
$\epsilon_{nl}$. The wavefunction can be written as $\Phi_{nl}({\bf
r})= \phi_{nl}(r) Y_{lm}(\theta,\phi)/r$ where $Y_{lm}(\theta,\phi)$
are the spherical harmonics. The orbitals $\phi_{nl}(r)$ satisfy the
radial Schr\"odinger equation:
\be
{\left[\hat{T}+\hat{V}_{\mathrm{ion}}+\hat{V}_{\mathrm{HF}}[\{\phi(r)\}]\right]}\phi_{nl}(r)=\epsilon_{nl} \phi_{nl}(r), \label{eq:shro}
\ee
where $\hat{T}=-d^2/(2dr^2) +l(l+1)/(2r^2)$, $\hat{V}_{\mathrm{ion}}(r)=-{Z}/{r}$ is
the bare nuclear potential, and $Z$ is the atomic number.
$\hat{V}_{\mathrm{HF}}[\{\phi(r)\}]$ is the HF potential
defined such that, 
\be
\hat{V}_{\rm{HF}}[\{\phi(r)\}] \phi_{nl}({ r}) =
\hat{V}_H[\{\phi(r)\}]\phi_{nl}({ r}) + \hat{V}_x^{nl}[\{\phi(r)\}]
\phi_{nl}({ r}); \label{eq:exch}
\ee
\be
\hat{V}_H[\{\phi(r)\}] = \sum_{nl} \int d{\bf r'} \,\frac{|\Phi_{nl}({\bf{r'}})|^2}{
  |{\bf r}  -\bf{r'}|}; \label{eq:hardef}
\ee
\be 
\hat{V}_{x}^{nl}[\{\phi(r)\}] \Phi_{nl}({\bf r}) =  \sum_{n'l'} \int
d{\bf r'}\,
\frac{\Phi_{nl}({\bf r'}) \Phi_{n'l'}^{*}({\bf{r'})}}{|{\bf r}  -\bf{r'}|} \Phi_{n'l'}({\bf{r}}). \label{eq:exchdef}
\ee
For convenience, equal and opposite self interaction terms are
included in both the Hartree and exchange terms.  Therefore, the
Hartree potential $\hat{V}_H[\{\phi(r)\}] \equiv \hat{V}_H[\rho(r)]$
is a functional of the spherically averaged total  electronic density $\rho(r)=
\sum_{nl}|\phi_{nl}(r)|^2$.

The second step in the construction of norm-conserving
pseudopotentials typically begins with the generation of a smooth set
of pseudo-orbitals $\tilde{\phi}_{nl}(r)$ to replace the AE orbitals
$\phi_{nl}(r)$, such that $\tilde{\phi}_{nl}(r)$ equals
the all-electron orbital beyond some cutoff distance
$r_c$. 
Different norm-conserving
pseudopotentials mainly differ in the form of $\tilde{\phi}_{nl}(r)$
for $r < r_c$, and the guiding principles used to
construct $\tilde{\phi}_{nl}(r)$.

In the RRKJ method, the pseudo-orbitals in the
core region are expressed as a linear combination of spherical Bessel
functions $j_l(q_k r)$
\be 
\tilde{\phi}_{nl}(r)=\left\{
\begin{aligned}
\sum_{k=1}^{N_b} c_{nlk} \,\, r j_l(q_k r) &&  r <  r_c \\
\phi_{nl}(r) &&  r \ge r_c  \label{eq:rrkj_orb}.
\end{aligned}
\right.
\ee
The Bessel wave-vectors $q_k$ are chosen such that, 
\be
 \frac{j_{l}^{'}(q_k r_c)}{j_l(q_k r_c)}
=\frac{\phi_{nl}^{'}(r_c)}{\phi_{nl}(r_c)}
\;\;\;\;\;\;\;\;\; k=1,\ldots,N_b \label{eq:wave_qc} 
\ee
where $j_{l}^{'}(q_k r)$  and $\phi_{nl}^{'}(r)$ are the derivatives
of $j_{l}(q_k r)$ and $\phi_{nl}(r)$ with
respect to $r$, respectively.  $N_b$ is the number of Bessel coefficients, $c_{nlk}$.

The key ingredient of the RRKJ method is the optimization of the
Bessel coefficients, $c_{nlk}$, to minimize the residual kinetic energy
defined as follows:
\be 
\begin{aligned}
\Delta T_{nl}(\{c_{nlk}\},q_c) = -\int_0^\infty d {\bf r} \,  \tilde{\Phi}_{nl}({\bf
  r})\nabla^2 \tilde{\Phi}_{nl}({\bf r}) \\
 - \int_0^{q_c}  d{\bf q}\, {\bf q}^2 |\tilde{\Phi}_{nl}({\bf q})|^2, \label{eq:resid_ke} 
\end{aligned}
\ee
where $\tilde{\Phi}_{nl}({\bf q})$ is the Fourier transform of real
space pseudo-orbital $\tilde{\Phi}_{nl}({\bf r}) =
\tilde{\phi}_{nl}(r) Y_{lm}(\theta,\phi)/r$, and $q_c$ is the target
wavevector above which the contribution to the residual kinetic energy
is as small as possible.  For a particular $q_c$, the planewave energy
cutoff required to achieve total energy convergence of $\Delta T_{nl}$
is given by the square of the maximum value of $q_c$ over all
orbitals \cite{RRKJ}.

The  optimization of $\Delta T_{nl}(\{c_{nlk}\},q_c)$ is constrained to
pseudo-orbitals that are continuous at
$r_c$, with continuous first and second derivatives [in fact, one of
these local constraints would be already satisfied due to
Eq.(\ref{eq:wave_qc})], and that satisfy norm conservation of the
core charge:
\be
\int_0^{r_c} dr\, |\tilde{\phi}_{nl}(r)|^2 = \int_0^{r_c}  dr \,|\phi_{nl}(r)|^2. \label{eq:normconv} 
\ee
This procedure defines all the Bessel coefficients in
Eq.~(\ref{eq:rrkj_orb}), and $N_b$ must be at least 3 to allow the
constraints to be satisfied.  Any additional parameters will be
optimized by minimizing the residual kinetic energy due to the
non-linearity of Eq.~(\ref{eq:resid_ke}). Throughout this study, we
used $N_b=6$.

The screened pseudopotential ${\hat{V}}_{\mathrm{scr}}^{nl}(r)$ is
defined as the potential that makes the desired pseudo-orbital
$\tilde{\phi}_{nl}(r)$ an eigenstate of the one-electron Schr\"odinger
equation, with the same eigenvalue $\epsilon_{nl}$ as the
corresponding all-electron state:

\be
\left[\hat{T}+{\hat{V}}_{\mathrm{scr}}^{nl}(r)\right]\tilde{\phi}_{nl}(r)=\epsilon_{nl}
\,\tilde{\phi}_{nl}(r). 
\ee 
This equation
can now be inverted, thanks to the nodeless character of
$\tilde{\phi}_{nl}(r)$, to obtain the screened pseudopotential:
\be
{{\hat{V}}_{\mathrm{scr}}}^{nl}(r)= \epsilon_{nl}(r)- \frac{\ell(\ell+1)}{2\,r^2} +
  \frac{1}{2{\tilde{\phi}_{nl}(r)}}
  \frac{d^2}{dr^2}\left[\tilde{\phi}_{nl}(r)\right]. 
\ee 
Note that ${\hat{V}}_{\mathrm{scr}}^{nl}(r)$ is a continuous function
because of the continuity requirements imposed on the pseudo-orbital
$\tilde{\phi}_{nl}(r)$  and its first two derivatives.

The last step in the construction of pseudopotentials is to remove the
screening effects of the valence electrons, and to obtain the ionic
pseudopotential. This is done by subtracting the Hartree and
exchange-potential contributions of the valence electrons from the
screened potential ${{\hat{V}}}_{\mathrm{scr}}^{nl}(r)$. That is,
\be
{{\hat{V}}}_{\mathrm{ion}}^{nl}(r)=
{{\hat{V}}}_{\mathrm{scr}}^{nl}(r)-\hat{V}_H[\{\tilde{\phi}(r)\}_{v}]-
    \frac{\hat{V}_{{x}}^{nl}[\{\tilde{\phi}(r)\}_{v}] \tilde{\phi}_{nl}(r)}{\tilde{\phi}_{nl}(r)},     \label{eq:vion}
\ee
where $\{\tilde{\phi}(r)\}_{v}$ includes only the valence
orbitals. Note that the last  term in Eq.~(\ref{eq:vion}), the
exchange potential, is orbital dependent, and its explicit construction
is feasible because the pseudo-orbitals are nodeless.

Up to this point, the pseudopotential construction recipe mirrors
exactly the procedure done in  regular norm-conserving
pseudopotentials based on DFT, except that the original Hamiltonian is
the HF and not the KS Hamiltonian. However, the resulting
ionic potentials, ${{\hat{V}}}_{\mathrm{ion}}^{nl}(r)$, have different
behavior for large $r$, which in turn stems from the corresponding
large-$r$ behavior of the HF and KS orbitals. It was shown by Handeler
\etal \cite{handler_limits_hf} that the HF orbitals decay at
infinity as
\be
\phi_{nl}(r)= r^{\beta_{nl}+1} e^{-\alpha r} \left[a_{nl}+\frac{b_{nl}}{r} +
  {\cal{O}}(1/r^2)\right], \label{eq:hflimit}
\ee
where $\alpha= (-2\epsilon_{\mathrm{HO}})^{1/2}$ is determined by the
eigenvalue of the highest occupied orbitals $\epsilon_{\mathrm{HO}}$, and
$\beta_{nl}$ is an orbital dependent quantity \cite{handler_limits_hf}. This
behavior is to be contrasted with the  exponential decaying
behavior of the KS orbitals, $\phi_{nl}(r) \propto
\exp[(-2\epsilon_{nl})^{1/2} r ]$ for large $r$.

In DFT-based pseudopotentials, the
exchange potential is replaced by the exchange-correlation
functional of the electron density $\rho(r)$, which decays
exponentially to zero for large $r$. Thus, the ionic potential
${{\hat{V}}}_{\mathrm{ion}}^{nl}(r)= -Z_{ v}/r$ for large
$r$ where $Z_{v}$ is the valence charge density. 

The effect of the special long-range behavior of the HF orbitals on
the ionic pseudopotential ${{\hat{V}}}_{\mathrm{ion}}^{nl}(r)$ can be
understood  by writing
${\hat{V}}_{\mathrm{ion}}^{nl}(r)$ as,
\be 
\begin{aligned}
 {\hat{V}}_{\mathrm{ion}}^{nl}(r)= - \frac{Z}{r}
+\hat{V}_H[\rho-\tilde{\rho_{v}}] +  \frac{\hat{V}_{{x}}^{nl}[\{{\phi}(r)\}]
 {\phi}_{nl}(r)}{{\phi}_{nl}(r)}  \\ 
 - \frac{\hat{V}_{{x}}^{nl}[\{\tilde{\phi}(r)\}_{v}]
 \tilde{\phi}_{nl}(r)}{\tilde{\phi}_{nl}(r)}\,\,\,\,\, {\rm for\,\,\,} r >
  r_c, \label{eq:vion1} 
\end{aligned}
\ee
where the Schr\"odinger equation of Eq.~(1) is used.  Here $\rho(r)$
is the total electronic charge density while $\tilde{\rho}_{v}(r)$ is
the pseudo-valence charge density. In the HF case, the exchange
potential is a non-local functional of the HF orbitals as shown in
Eq.~(\ref{eq:exchdef}). In particular, the leading behavior of the
numerator for large $r$ will be $r^{\beta} e^{-\alpha r}$ where
$\beta=\max_{nl}\beta_{nl}$,
while the denominator will go as
$r^{\beta_{nl}+1} e^{-\alpha r}$.
Therefore, the exchange potential, as computed in Eq.~(\ref{eq:vion1}),
is not guaranteed to  decay faster than 
$1/r$. In fact, 
the exchange potential can even grow without bound for large $r$
when computed in this way.
This is the source of the non-local long-range non-Coulombic tail in
the HF pseudopotentials \cite{trail_needs1}.

\begin{table}[pt]
\caption{
\label{tab:tabrc}
A summary of the cutoff radii (in atomic units) for the
pseudopotentials for $s$, $p$, and $d$ channels used in the atomic
calculations.  The last column shows $q_{c}^{\rm{max}}$, the largest
$q_c$ value over all the orbitals (in Ry$^{1/2}$). The cutoff energy for
a planewave basis is approximately the square of $q_{c}^{\rm{max}}$.}
\begin{ruledtabular}
\begin{tabular}{lccccc}
\multicolumn{1}{r}{Atom} & $r_{cs}$ &$r_{cp}$ & $r_{cd}$ & $q_{c}^{\rm{max}}$ \\ 
\hline
H       & 0.50 & 0.50 & 0.50 & 10.8 \\
He      & 0.60 & 0.60 & 0.60 & 12.8 \\
Li      & 2.19 & 2.37 & 2.37 & 2.7 \\
Be      & 1.41 & 1.41 & 1.41 & 6.2 \\
B       & 1.88 & 1.96 & 1.96 & 4.0 \\
C       & 1.10 & 1.10 & 1.10 & 8.3 \\
N       & 0.94 & 0.88 & 0.84 & 10.6 \\
O       & 0.80 & 0.75 & 0.99 &  12.6 \\
F       & 0.70 & 0.64 & 0.89 &  15.0 \\
Ne      & 0.63 & 0.57 & 0.63 &  17.1 \\
Na      & 2.70 & 2.85 & 2.85 &  2.5 \\
Mg      & 2.38 & 2.38 & 2.38 &  3.2 \\
Al      & 1.94 & 2.28 & 2.28 &  3.9 \\
Si      & 1.67 & 2.01 & 2.06 &  4.6 \\
P       & 1.48 & 1.71 & 1.71 &  5.8 \\
S       & 1.33 & 1.50 & 1.50 &  5.8 \\
Cl      & 1.19 & 1.34 & 1.34 &  6.5 \\
Ar      & 1.09 & 1.20 & 1.31 &  7.3 \\
\end{tabular}
\end{ruledtabular}
\end{table}

\begin{figure}[t]
\includegraphics[width=3.5in,clip]{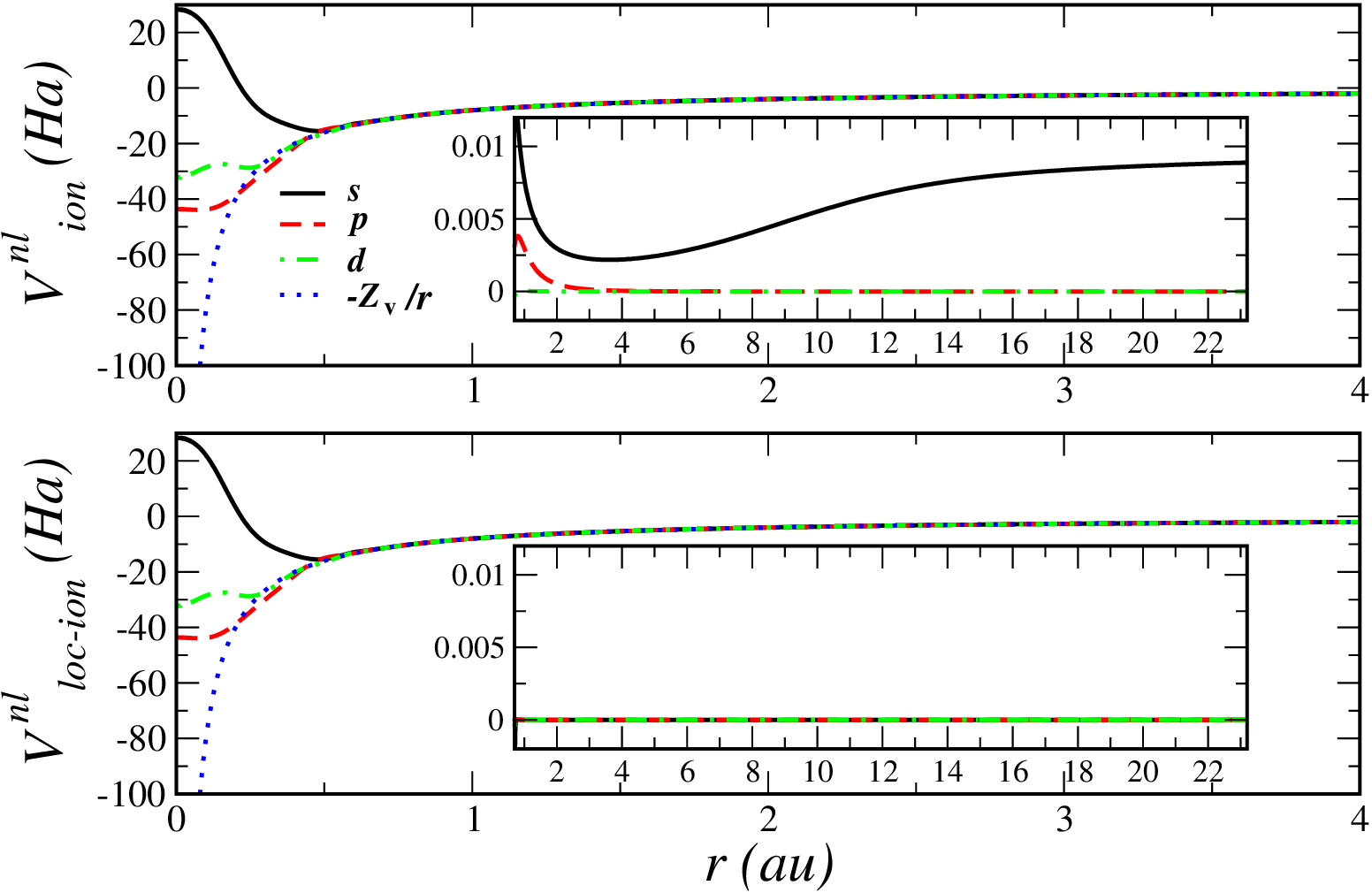}
\caption{\label{fig:nepot} (Color online) The upper panel shows the Ne
  potential $V_{\rm{ion}}^{nl}$ constructed using the RRKJ method
  before the tail correction. The inset shows $\Delta V_{\rm{ion}} =
  V_{\rm{ion}}^{nl}+ Z_v/r $ which displays the non-physical tail of
  the $s$-potential channel for large $r$.  The lower panel shows
  $V_{\rm{loc-ion}}^{nl}$ after the tail correction procedure. In the
  inset, we show the large-$r$ behavior of $\Delta V_{\rm{loc-ion}} =
  V_{\rm{loc-ion}}^{nl}+Z_v/{r}$.}

\end{figure}

\begin{figure}[t]
\includegraphics[width=3.5in,clip]{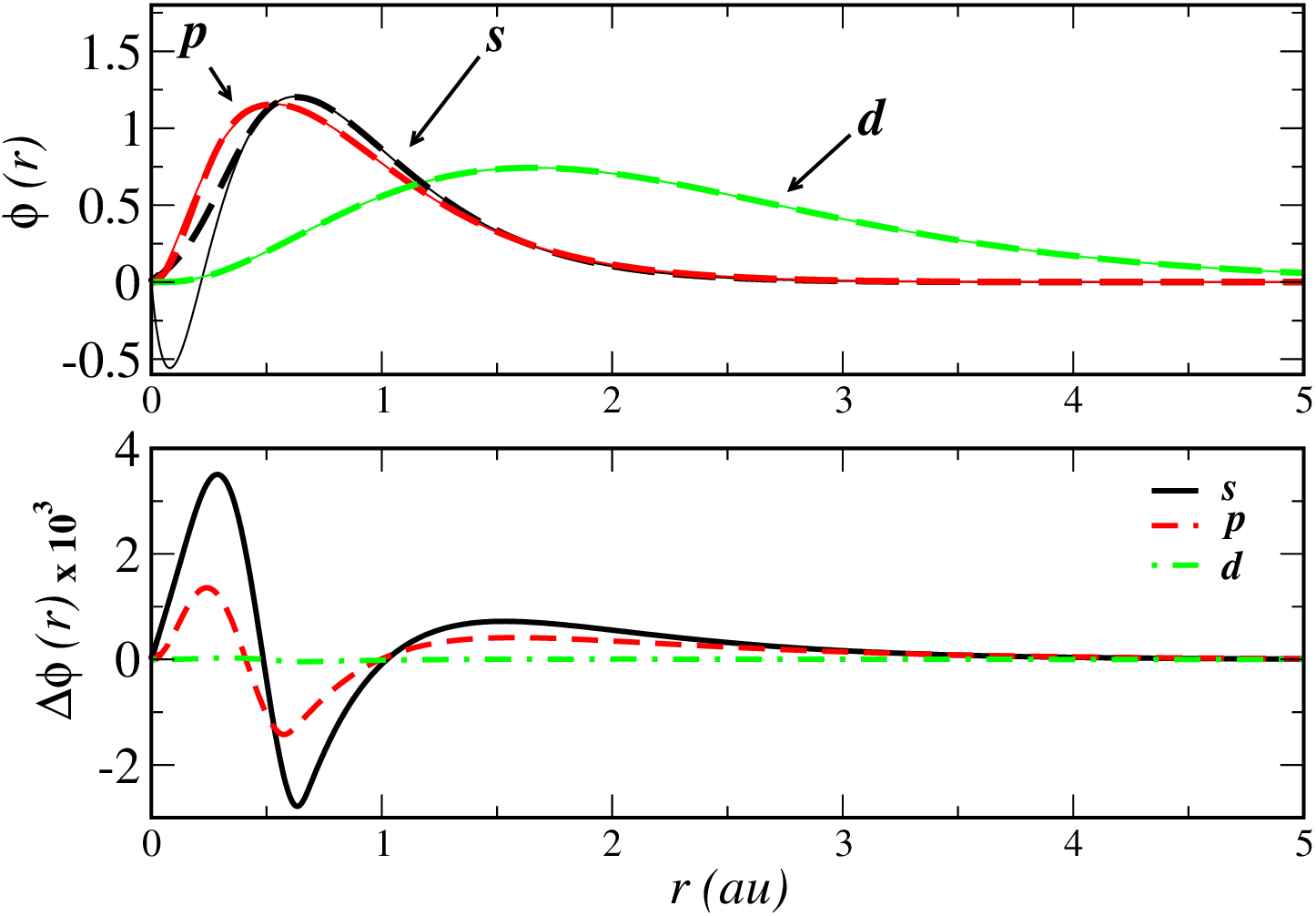}
\caption{\label{fig:nepsi} (Color online) The upper panel shows the
  all-electron (solid lines) and the localized pseudo-orbitals (dashed
  lines) of Ne constructed using the RRKJ method. The lower panel
  shows the difference in the pseudo-orbitals before and after the
  localization procedure (scaled by 10$^{-3}$).}
\end{figure}

In this work we follow the Trail and Needs self-consistent approach to
localize the ionic pseudopotential \cite{trail_needs1}. The localized
ionic potential ${\hat{V}}^{nl}_{{\rm{loc-ion}}}(r)$, which is forced
to behave as $-Z_{ v}/r$ for large $r$, is defined as 
\be
{\hat{V}}^{nl}_{{\rm{loc-ion}}}(r) =  \left\{
\begin{array}{ll}
\gamma_{nl}(r) +\hat{V}^{nl}_{{\rm{\mathrm{ion}}}}(r)   &   r < r_{\mathrm{loc}} \\\\\\
e^{-\xi(r-r_{\mathrm{loc}})^2} [ \gamma_{nl}(r)
  +\hat{V}^{nl}_{{\rm{\mathrm{ion}}}}(r)+\\\\
Z/r- \hat{V}_{\rm{H}}(\rho_{\rm{c}}) ] + \hat{V}_{\rm{H}}(\rho_{\rm{c}}) - Z/{r}  &     r \ge
r_{\mathrm{loc}} 
\label{eq:loc_v}
\end{array}
\right.
\ee
where $\rho_{\rm{c}}(r)$ is the core charge density,
$1/\sqrt{\xi}=r_{\rm{loc}}/16$ is the characteristic distance for the
localized ${\hat{V}}^{nl}_{{\rm{loc-ion}}}(r)$ to approach $-Z_{
v}/r$, and $r_{\mathrm{loc}}$ is the localization radius.  For all
pseudopotentials, the value of $r_{\mathrm{loc}}$ for each $l$-channel
is fixed to be the same as the corresponding $r_c$.  $\gamma_{nl}(r)$
is defined as $p_{nl}+ q_{nl} f(r,r_{\mathrm{loc}})$ where $p_{nl}$
and $q_{nl}$ are parameters which are chosen to localize the ionic
pseudopotential ${{\hat{V}}}_{\mathrm{ion}}^{nl}(r)$ in a
self-consistent manner, and $f(r,r_{\mathrm{loc}})$ is a polynomial
function whose explicit form depends on the pseudopotential
construction method (see below).

The self-consistent procedure is performed as follows.
First, starting with an initial guess for the $p_{nl}$ and $q_{nl}$
parameters, the HF equations with a new ionic potential, as defined in
Eq.~(\ref{eq:loc_v}), are solved for a new set of orbitals and
eigenvalues. The $p_{nl}$ and $q_{nl}$ parameters are then adjusted to
minimize the error in the eigenvalues and logarithmic
derivatives\cite{logdernote1}.  The procedure is then repeated until
$p_{nl}$ and $q_{nl}$ parameters are found that give eigenvalues and
logarithmic derivatives that closely match the all-electron values to
the desired tolerance.

Finally, we define $f(r,r_{\mathrm{loc}})$ used in
$\gamma_{nl}(r)$. For the case of Troullier-Martins pseudopotentials,
we chose the same form as was used in Ref. \cite{trail_needs1}, namely
$f(r,r_{\rm{loc}})=r^4\left(1-{2\,r^2}/{3
r_{\mathrm{loc}}^{2}}\right)$ for $r < r_{\rm{loc}}$, and
$f(r,r_{\rm{loc}})=r_{\mathrm{loc}}^{4}/{3}$ otherwise.  For RRKJ
pseudopotentials, we investigated a few different forms and found the
results to be insensitive to the choices. We will report our results
using
 $f(r,r_{\rm{loc}})=1-r/2r_{\mathrm{loc}}$ for $r < r_{\rm{loc}}$, and
$f(r,r_{\rm{loc}})={r_{\mathrm{loc}}}/{2}$ for $r \ge
  r_{\rm{loc}}$.

\begin{figure}[tbh]
\includegraphics[width=3.5in,clip]{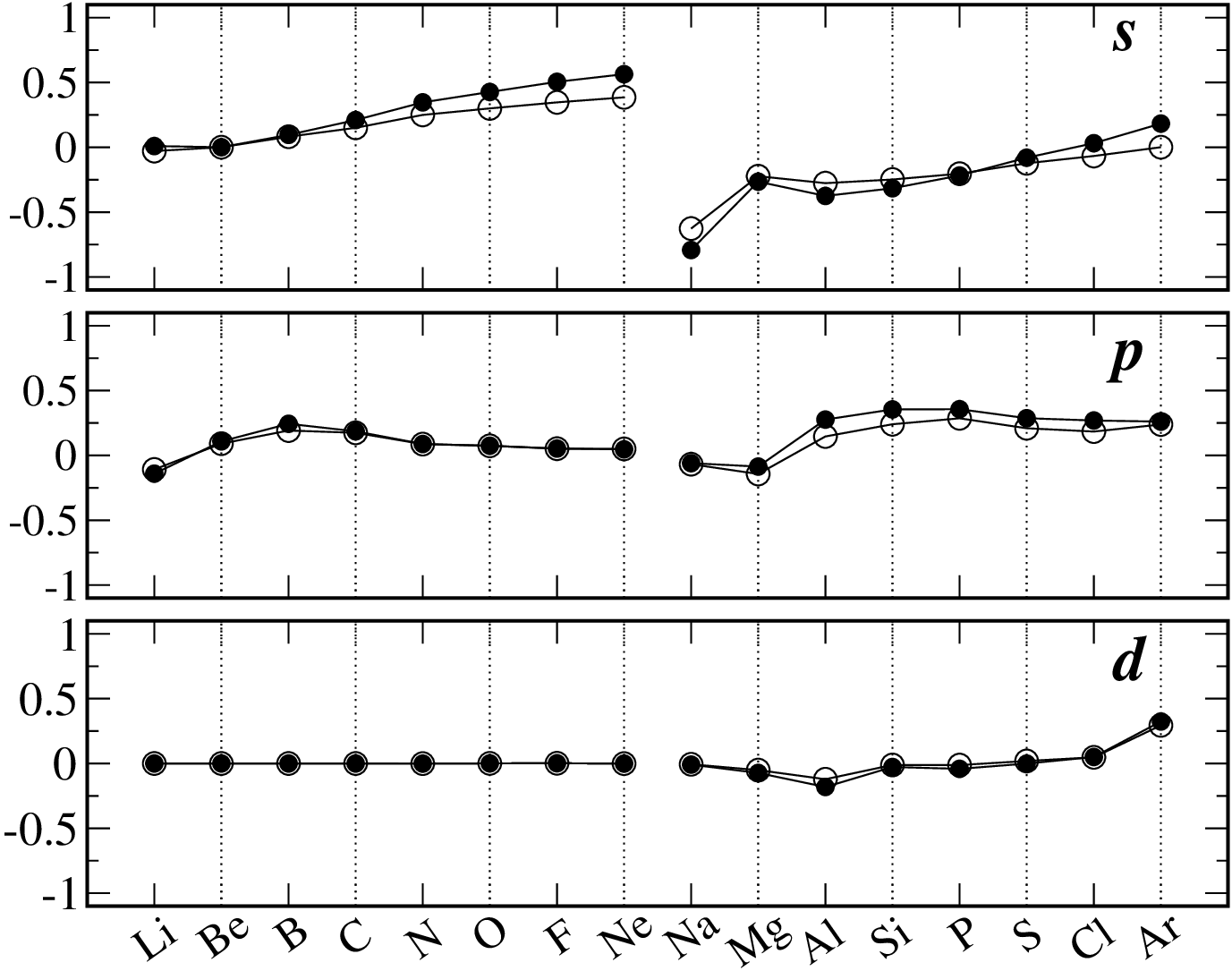}
\caption{
  \label{fig:norm_error} Amount of the violation of the
  core norm-conservation for the $s$, $p$, and $d$ pseudo-orbitals (in
  units of $10^{-3}$~electrons), respectively. There is less than $0.001$
  electron shift between the valence and core
  regions. Pseudopotentials are constructed using the RRKJ (open
  circles), and TM (closed circles) methods.}
\end{figure}

\begin{figure}[tbh]
\includegraphics[width=3.5in,clip]{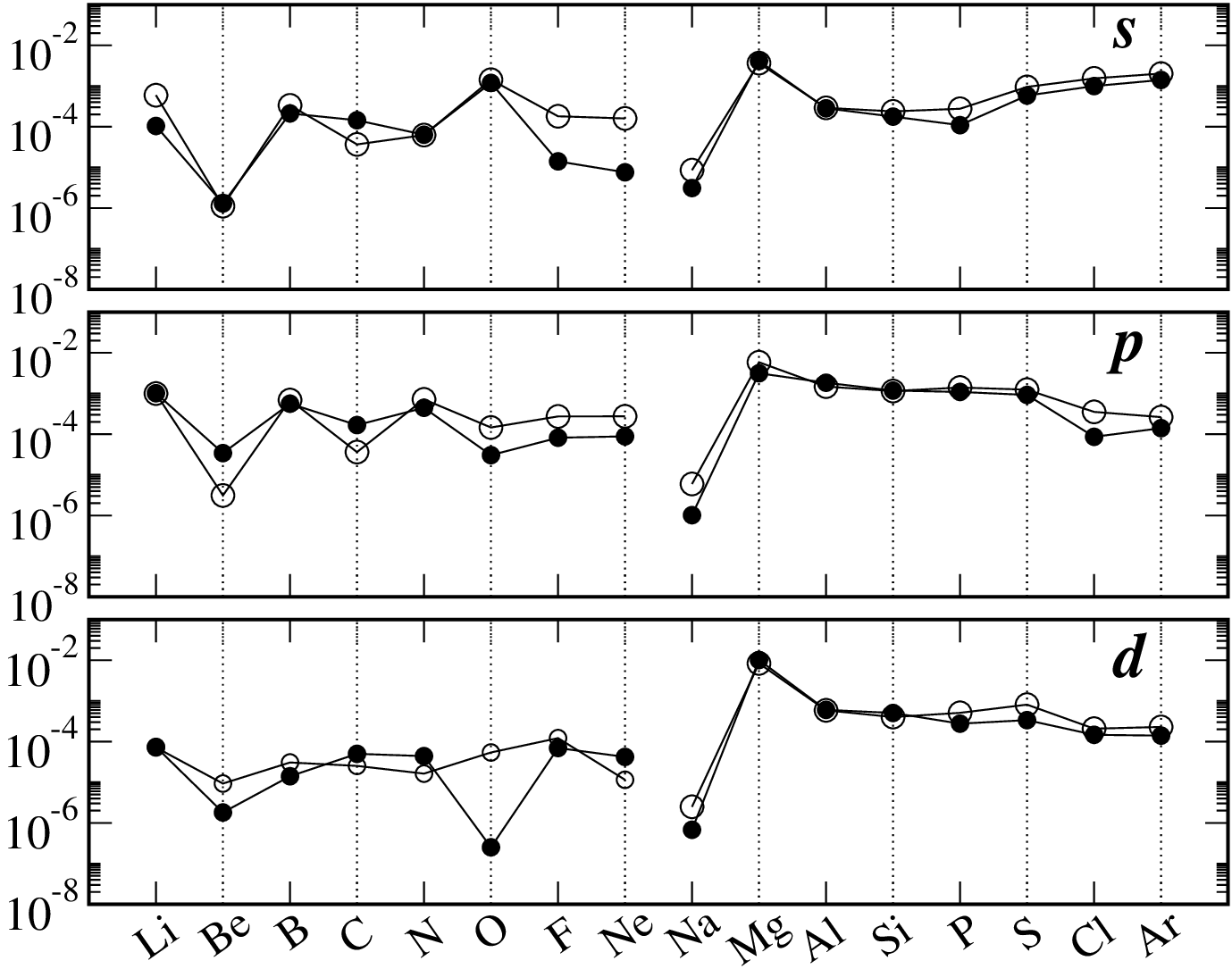}
\caption{ \label{fig:logder_error} Similar to Fig.~3, except that we
show the relative error in the logarithmic derivative at $r_c$ from
the all-electron values. The relative error is defined as
$|{1-L_{loc}}/{L}|$, where $L$ and  $L_{loc}$ are the logarithmic
derivative before and after the localization procedure is applied,
respectively.  The logarithmic derivative is approximately conserved
in the localization procedure.}
\end{figure}

\begin{table*}[tbh]
\caption{
\label{tab:tabion}
Ionization energy study of the first- and second-row elements. The
  first column shows the all-electron ionization energy, while the
  other columns show the deviation from the AE result. Columns 2 and 3
  show the pseudopotential ionization energies as obtained using the
  original and the localized RRKJ pseudopotentials
  [Eqs.~(\ref{eq:vion1}) and (\ref{eq:loc_v}), respectively]. Columns
  4 and 5 show the same values as obtained using the TM method. For
  comparison, we show also the values obtained by Trail and Needs
  \cite{trail_needs1}. The pseudopotential parameters are summarized
  in Table~\ref{tab:tabrc}. We show also the average abs. error for
  each pseudopotential with respect to the AE values.  All energies are in Hartrees.}
\begin{ruledtabular}
\begin{tabular}{lcdddddddd} 
& \multicolumn{1}{c}{AE}&\multicolumn{2}{c}{This work/RRKJ} & & \multicolumn{2}{c}{This work/TM} & & \multicolumn{2}{c}{TM\footnotemark[1]} \\
Atom &  $E_{\rm ion}$  & \multicolumn{1}{c}{$\Delta{E}_{{\rm ion}}$} & \multicolumn{1}{c}{$\Delta\tilde{E}_{{\rm ion}}$} & & \multicolumn{1}{c}{$\Delta{E}_{{\rm ion}}$} & \multicolumn{1}{c}{$\Delta\tilde{E}_{{\rm ion}}$} & & \multicolumn{1}{c}{$\Delta{E}_{{\rm ion}}$} & \multicolumn{1}{c}{$\Delta\tilde{E}_{{\rm ion}}$}  \\ \hline
H       &       0.5000 &       0.0000     &       0.0000     &               &       0.0000     &       0.0000     &               &       0.0000     &       0.0000     \\
He      &       0.8617 &       -0.0009    &       -0.0008    &               &       -0.0007    &       -0.0007    &               &       -0.0006    &       -0.0006    \\
Li      &       0.1963 &       0.0000     &       0.0000     &               &       0.0000     &       0.0000     &               &       0.0000     &       0.0000     \\
Be      &       0.2956 &       -0.0002    &       -0.0002    &               &       -0.0002    &       -0.0002    &               &       -0.0002    &       -0.0002    \\
B       &       0.2915 &       -0.0006    &       -0.0006    &               &       -0.0005    &       -0.0005    &               &       -0.0005    &       -0.0005    \\
C       &       0.3964 &       -0.0009    &       -0.0008    &               &       -0.0006    &       -0.0005    &               &       -0.0006    &       -0.0005    \\
N       &       0.5129 &       -0.0009    &       -0.0008    &               &       -0.0006    &       -0.0005    &               &       -0.0006    &       -0.0006    \\
O       &       0.4368 &       0.0001     &       0.0001     &               &       0.0001     &       0.0001     &               &       0.0001     &       0.0000     \\
F       &       0.5776 &       -0.0004    &       -0.0003    &               &       -0.0003    &       -0.0002    &               &       -0.0003    &       -0.0003    \\
Ne      &       0.7293 &       -0.0008    &       -0.0007    &               &       -0.0006    &       -0.0004    &               &       -0.0005    &       -0.0005    \\
Na      &       0.1819 &       -0.0002    &       -0.0002    &               &       -0.0001    &       -0.0001    &               &       -0.0001    &       -0.0001    \\
Mg      &       0.2428 &       -0.0002    &       -0.0002    &               &       -0.0002    &       -0.0002    &               &       -0.0002    &       -0.0002    \\
Al      &       0.2020 &       -0.0003    &       -0.0003    &               &       -0.0002    &       -0.0002    &               &       -0.0002    &       -0.0002    \\
Si      &       0.2812 &       -0.0005    &       -0.0005    &               &       -0.0005    &       -0.0004    &               &       -0.0005    &       -0.0005    \\
P       &       0.3690 &       -0.0009    &       -0.0008    &               &       -0.0008    &       -0.0008    &               &       -0.0008    &       -0.0008    \\
S       &       0.3317 &       0.0011     &       0.0011     &               &       0.0012     &       0.0012     &               &       0.0011     &       0.0011     \\
Cl      &       0.4335 &       -0.0003    &       -0.0002    &               &       0.0000     &       0.0000     &               &       -0.0002    &       -0.0002    \\
Ar      &       0.5430 &       -0.0015    &       -0.0014    &               &       -0.0015    &       -0.0014    &               &       -0.0015    &       -0.0014    \\
\\
\\
\multicolumn{2}{c}{Average abs. error}                       &       0.0005     &       0.0005     &               &       0.0004     &       0.0004     &               &       0.0005     &       0.0004     \\
\multicolumn{2}{c}{Maximum abs. error}                       &       0.0015     &       0.0014     &               &       0.0015     &       0.0014     &               &       0.0015     &       0.0014     \\
\end{tabular}
\end{ruledtabular}
\footnotetext[1]{Ref.~\cite{trail_needs1}.}
\end{table*}

\begin{table*}[tbh]
\caption{
\label{tab:tab_EA} Similar to Table~\ref{tab:tabion}, except we show
the electron affinity of the first- and second-row elements. }
\begin{ruledtabular}
\begin{tabular}{ldddcdd} 
& \multicolumn{1}{c}{AE}&\multicolumn{2}{c}{RRKJ} & & \multicolumn{2}{c}{TM}  \\
Atom & \multicolumn{1}{c}{$E_{\rm EA}$}  & \multicolumn{1}{c}{$\Delta{E}_{{\rm EA}}$} & \multicolumn{1}{c}{$\Delta\tilde{E}_{{\rm EA}}$} & & \multicolumn{1}{c}{$\Delta{E}_{{\rm EA}}$} & \multicolumn{1}{c}{$\Delta\tilde{E}_{{\rm EA}}$} \\ \hline
B       &       -0.0098        &       0.0001     &       0.0002     & &     0.0     &       0.0001     \\
C       &       0.0202         &       0.0002     &       0.0004     & &     0.0001     &       0.0002     \\
N       &       -0.0790        &      -0.0004    &       -0.0001    & &     -0.0003    &       0.0000    \\
O       &       -0.0196        &       0.0000     &       0.0003     & &     -0.0001    &       0.0002     \\
F       &       0.0501         &       0.0002     &       0.0006     & &     0.0001     &       0.0004     \\
Al      &       0.0016         &       0.0002     &       0.0002     & &     0.0002     &       0.0002     \\
Si      &       0.0353         &       0.0005     &       0.0005     & &     0.0005     &       0.0005     \\
P       &       -0.0199        &      -0.0009    &       -0.0007    & &     -0.0008    &       -0.0007    \\
S       &       0.0335         &       0.0001     &       0.0002     & &     0.0000     &       0.0002     \\
Cl      &       0.0948         &       0.0011     &       0.0013     & &     0.0010     &       0.0011     \\
\\
\\
\multicolumn{2}{c}{Average abs. error}                &       0.0004     &       0.0004     & &     0.0003     &       0.0004     \\
\multicolumn{2}{c}{Maximum abs. error}                     &       0.0011     &       0.0013     & &     0.0010     &       0.0011     \\
\end{tabular}
\end{ruledtabular}
\end{table*}

To illustrate the pseudopotential localization procedure, we show in
the upper panel of Fig.~\ref{fig:nepot} the pseudopotentials
${\hat{V}}^{nl}_{{\rm{ion}}}(r)$ of Ne as constructed using the RRKJ
method with the parameters given in Table~I.  In the lower panel, we
show the same Ne pseudopotentials after the self-consistent
localization procedure. As can be seen, the differences between the
upper and lower panels are rather small, and are in the core as well
as in the valence regions. Most of the changes are within the core
region, however. The insets show the long-range behavior of
${\hat{V}}^{nl}_{{\rm{ion}}}(r) + Z_v/r$. Note, in particular, in the
upper panel, the non-decaying tail of the $s$-potential, which can
be shown \cite{trail_needs1} to behave as $a+b/r$ for large $r$ where
$a$ and $b$ are real numbers. 

In Fig.~\ref{fig:nepsi}, we show in the upper panel the all-electron
and pseudo-orbitals after the localization procedure.  The lower panel
shows the difference in the pseudo-orbitals before and after the
localization procedure. Again, changes are found in both the core and
valence regions. One obvious drawback of the localization procedure is
that the core-norm will be different from the all-electron value, {\rm
{i.e.}}~Eq.~(\ref{eq:normconv}) will be violated slightly.  This can
be seen in Fig.~\ref{fig:nepsi} because the changes in the orbitals
for $r<r_c$ are all positive, and will integrate to a non-zero value.

As mentioned before, the localization procedure conserves the
all-electron eigenvalues, and the logarithmic derivative. In practice,
the logarithmic derivative is only conserved approximately. Figures
\ref{fig:norm_error} and \ref{fig:logder_error} show a systematic
study of the violations of the norm-conservation and the change in the
logarithmic derivative at $r_c$ due to the localization procedure for
the first- and second-row elements. The study was done using both the
RRKJ and TM pseudopotentials. From Fig.~\ref{fig:norm_error}, we see
that there is less than $\approx 0.001$~electron shift between the
core and the valence. 
The relative error between the logarithmic derivatives before and
after applying the localization procedure is shown in
Fig.~\ref{fig:logder_error}~\cite{logdernote2}. As can be seen from both
figures, both pseudopotential construction methods yield similar
results.


\begin{table*}[ptbh]
\caption{\label{tab:tabexc} Excitation energy study of the first- and 
second-row elements. The first column shows the ground
state configuration, and the second column shows the configuration of
the excited states. The all-electron excitation energy is shown in the
third column. The other columns show the deviation from
the AE result, similar to
Table~\ref{tab:tabion}. All energies are in Hartrees. } 
\begin{ruledtabular}
\begin{tabular}{llcdddddddd}
\multicolumn{1}{c}{Atom}  &  & \multicolumn{1}{c}{AE} &\multicolumn{2}{c}{RRKJ} &&\multicolumn{2}{c}{TM} &&\multicolumn{2}{c}{TM\footnotemark[1]}  \\
\multicolumn{1}{c}{Ground state}   & \multicolumn{1}{c}{Excited state} & \multicolumn{1}{c}{$E_{{\rm exc}}$} & \multicolumn{1}{c}{$\Delta{E}_{{\rm exc}}$} & \multicolumn{1}{c}{$\Delta\tilde{E}_{{\rm exc}}$}  && \multicolumn{1}{c}{$\Delta{E}_{{\rm exc}}$} & \multicolumn{1}{c}{$\Delta\tilde{E}_{{\rm exc}}$} & &\multicolumn{1}{c}{$\Delta{E}_{{\rm exc}}$} & \multicolumn{1}{c}{$\Delta\tilde{E}_{{\rm exc}}$} \\ \hline
H $1s^1 [^2S]$  &       $2p^1 [^2P]$                    &       0.3750  &       0.0000     &       0.0000     & &     0.0000     &       0.0000     & &     0.0000     &       0.0000     \\
                &       $3d^1 [^2D]$                    &       0.4444  &       0.0000     &       0.0000     & &     0.0000     &       0.0000     & &     0.0000     &       0.0000     \\
\\
He $1s^2 [^1S]$ &       $1s^1 2p^1 [^3P]$               &       0.7302  &       -0.0008    &       -0.0008    & &     -0.0006    &       -0.0006    & &     -0.0006    &       -0.0006    \\
                &       $1s^1 3d^1 [^3D]$               &       0.8061  &       -0.0009    &       -0.0008    & &     -0.0007    &       -0.0007    & &     -0.0006    &       -0.0006    \\
\\
Li $2s^1 [^2S]$ &       $2p^1 [^2P]$                    &       0.0677  &       0.0000     &       0.0000     & &     0.0000     &       0.0000     & &     0.0000     &       0.0000     \\
                &       $3d^1 [^2D]$                    &       0.1408  &       0.0000     &       0.0000     & &     0.0000     &       0.0000     & &     0.0000     &       0.0000     \\
\\
Be $2s^2 [^1S]$ &       $2s^1 2p^1 [^3P]$               &       0.0615  &       0.0011     &       0.0011     & &     0.0010     &       0.0010     & &     0.0020     &       0.0020     \\
                &       $2s^1 3d^1 [^3D]$               &       0.2389  &       -0.0002    &       -0.0002    & &     -0.0002    &       -0.0002    & &     -0.0001    &       -0.0001    \\
\\
B $2s^2 2p^1 [^2P]$     &       $2s^1 2p^2 [^4P]$       &       0.0784  &       0.0023     &       0.0022     & &     0.0020     &       0.0019     & &     0.0020     &       0.0019     \\
        &       $2s^2 3d^1 [^2D]$                       &       0.2353  &       -0.0006    &       -0.0006    & &     -0.0005    &       -0.0005    & &     -0.0005    &       -0.0005    \\
\\
C $2s^2 2p^2 [^3P]$     &       $2s^1 2p^3 [^5S]$       &       0.0894  &       0.0060     &       0.0058     & &     0.0054     &       0.0053     & &     0.0054     &       0.0053     \\
        &       $2s^2 2p^1 3d^1 [^3F]$                  &       0.3402  &       -0.0008    &       -0.0008    & &     -0.0006    &       -0.0005    & &     -0.0006    &       -0.0005    \\
\\
N $2s^2 2p^3 [^4S]$     &       $2s^1 2p^4 [^4P]$       &       0.4127  &       0.0029     &       0.0030     & &     0.0030     &       0.0030     & &     0.0031     &       0.0031     \\
        &       $2s^2 2p^2 3d^1 [^4F]$                  &       0.4565  &       -0.0009    &       -0.0008    & &     -0.0006    &       -0.0006    & &     -0.0005    &       -0.0005    \\
\\
O $2s^2 2p^4 [^3P]$     &       $2s^2 2p^3 3d^1 [^5D]$  &       0.3809  &       0.0001     &       0.0001     & &     0.0001     &       0.0001     & &     0.0000     &       0.0000     \\
        &       $2s^1 2p^5 [^3P] $                      &       0.6255  &       -0.0003    &       -0.0002    & &     -0.0002    &       -0.0001    & &     -0.0002    &       -0.0001    \\
\\
F $2s^2 2p^5 [^2P]$     &       $2s^2 2p^4 3d^1 [^4F]$  &       0.5220  &       -0.0004    &       -0.0003    & &     -0.0003    &       -0.0002    & &     -0.0003    &       -0.0002    \\
        &       $2s^1 2p^6 [^2S] $                      &       0.8781  &       -0.0052    &       -0.0051    & &     -0.0051    &       -0.0049    & &     -0.0051    &       -0.0049    \\
\\
Ne $2s^2 2p^6 [^1S]$    &       $2s^2 2p^5 3d^1 [^3F]$  &       0.6734  &       -0.0008    &       -0.0007    & &     -0.0005    &       -0.0004    & &     -0.0006    &       -0.0004    \\
        &       $2s^1 2p^6 3d^1 [^3D]$                  &       1.7565  &       -0.0051    &       -0.0049    & &     -0.0047    &       -0.0045    & &     -0.0048    &       -0.0045    \\
\\
Na $3s^1 [^2S]$ &       $3p^1 [^2P]$                    &       0.0725  &       -0.0001    &       -0.0001    & &     -0.0001    &       -0.0001    & &     -0.0002    &       -0.0002    \\
        &       $3d^1 [^2D]$                            &       0.1263  &       -0.0002    &       -0.0002    & &     -0.0001    &       -0.0001    & &     -0.0001    &       -0.0001    \\
\\
Mg $3s^2 [^1S]$ &       $2s^1 2p^1 [^3P]$               &       0.0679  &       0.0008     &       0.0008     & &     0.0007     &       0.0007     & &     0.0007     &       0.0008     \\
        &       $2s^1 3d^1 [^3D]$                       &       0.1843  &       -0.0002    &       -0.0002    & &     -0.0002    &       -0.0002    & &     -0.0002    &       -0.0002    \\
\\
Al $3s^2 3p^1 [^2P]$    &       $3s^1 3p^2 [^4P]$       &       0.0858  &       0.0008     &       0.0007     & &     0.0007     &       0.0007     & &     0.0008     &       0.0007     \\
        &       $3s^2 3d^1 [^2D]$                       &       0.1441  &       -0.0002    &       -0.0002    & &     -0.0002    &       -0.0002    & &     -0.0002    &       -0.0002    \\
\\
Si $3s^2 3p^2 [^3P]$    &       $3s^1 3p^3 [^5S]$       &       0.0913  &       0.0023     &       0.0022     & &     0.0023     &       0.0022     & &     0.0023     &       0.0022     \\
        &       $3s^2 3p^1 3d^1 [^3F]$                  &       0.2146  &       -0.0001    &       -0.0001    & &     -0.0001    &       -0.0001    & &     -0.0001    &       -0.0001    \\
\\
P $3s^2 3p^3 [^4S]$     &       $3s^1 3p^4 [^4P]$       &       0.3006  &       -0.0004    &       -0.0003    & &     -0.0003    &       -0.0003    & &     -0.0004    &       -0.0003    \\
        &       $3s^2 3p^2 3d^1 [^4F]$                  &       0.3023  &       -0.0008    &       -0.0007    & &     -0.0007    &       -0.0007    & &     -0.0007    &       -0.0007    \\
\\
S $3s^2 3p^4 [^3P]$     &       $3s^2 3p^3 3d^1 [^5D]$  &       0.2672  &       0.0014     &       0.0014     & &     0.0015     &       0.0015     & &     0.0013     &       0.0015     \\
        &       $3s^1 3p^5 [^3P] $                      &       0.4260  &       -0.0010    &       -0.0010    & &     -0.0010    &       -0.0010    & &     -0.0009    &       -0.0009    \\
\\
Cl $3s^2 3p^5 [^2P]$    &       $3s^2 3p^4 3d^1 [^4F]$  &       0.3733  &       -0.0001    &       -0.0001    & &     0.0001     &       0.0002     & &     -0.0002    &       -0.0001    \\
        &       $3s^1 3p^6 [^2S] $                      &       0.5653  &       -0.0020    &       -0.0020    & &     -0.0020    &       -0.0021    & &     -0.0018    &       -0.0018    \\
\\
Ar $3s^2 3p^6 [^1S]$    &       $3s^2 3p^5 3d^1 [^3F]$  &       0.4824  &       -0.0014    &       -0.0013    & &     -0.0014    &       -0.0012    & &     -0.0014    &       -0.0013    \\
        &       $3s^1 3p^6 3d^1 [^3D]$                  &       1.1597  &       -0.0028    &       -0.0027    & &     -0.0026    &       -0.0024    & &     -0.0026    &       -0.0024    \\
\\
\\
\multicolumn{3}{c}{Average abs. error}                                  &       0.0012     &       0.0011     & &     0.0011     &       0.0011     & &     0.0011     &       0.0011     \\
\multicolumn{3}{c}{Maximum abs. error}                                       &       0.0060     &       0.0058     & &     0.0054     &       0.0053     & &     0.0054     &       0.0053     \\

\end{tabular}
\end{ruledtabular}
\footnotetext[1]{Ref.~\cite{trail_needs1}.}
\end{table*}

\section{Pseudopotential Tests on atomic properties}

In this section, we study the transferability of the pseudopotentials
by looking at several atomic properties namely ionization energies,
electron affinities, and excitation energies of the first- and
second-row elements. All of these calculations are performed by
solving the HF equations in real space with the same code \cite{opium}
which is used to generate the pseudopotentials.

We show results obtained using HF pseudopotentials
constructed using both the RRKJ and TM methods. For the TM
pseudopotentials, we used the same code \cite{opium} which is used to
generate the RRKJ pseudopotentials but following the Troullier-Martins
construction scheme.

The parameters for all pseudopotentials used in the study of the
atomic properties are summarized in Table~\ref{tab:tabrc}. In order to
aid in comparison to previous results, the reference configurations
and the construction parameters are the same as of
Ref.~\cite{trail_needs1}.  In general, however, one has to choose
these parameters according to the targeted applications.  Moreover,
unlike the TM method, the RRKJ method has two additional adjustable
parameters for each pseudo-orbital, namely $N_b$ and $q_c$.  We set
$N_b=6$ for all pseudopotentials, and selected the $q_c$'s such that
$\Delta T_{nl} \approx 5$~meV/electron for each orbital.  As
mentioned before, the energy cutoff required to achieve this level of
energy convergence in the target calculations is approximately the
square of the largest $q_c$ value used.

In Table~\ref{tab:tabion} we present the ionization energies. For each
element, we calculated the ionization energy using our RRKJ and TM
pseudopotentials.  These values are compared with the all-electron
value (shown in the first column), and we only report the difference,
$\Delta E_{{\rm ion}}$, from the all-electron energy.
$\Delta{E}_{{\rm ion}}$ is obtained using the original pseudopotential
of Eq.~(\ref{eq:vion1}) with the unphysical tail behavior,
while $\Delta\tilde{E}_{{\rm ion}}$ is calculated using the localized
HF pseudopotential of Eq.~(\ref{eq:loc_v}). We show also for
comparison the same results as obtained using the HF TM
pseudopotentials as reported in Ref.~\cite{trail_needs1}.

For both pseudopotential construction schemes, 
the agreement between $\Delta{E}_{{\rm ion}}$ and
$\Delta\tilde{E}_{{\rm ion}}$ indicates that the
self-consistent procedure that is used to localize the pseudopotential
did not change significantly the original potential. It is important
to stress that the effects of the non-Coulombic tail in atomic
calculations are minimal, and this is why there is a good agreement
between $\Delta{E}_{{\rm ion}}$ and $\Delta\tilde{E}_{{\rm
ion}}$. However, in any solid calculation with periodic
boundary conditions, the results obtained using 
pseudopotentials with the tail problem would be erroneous. 

Also, the pseudopotential results in Table~\ref{tab:tabion} are in
good agreement with the all-electron values for both the RRKJ and TM
methods. On average the difference is less than $0.5$~milli-Hartree (mHa)
and the largest deviation is less than $1.5$~mHa. This is a
clear indication of the good quality of the pseudopotentials.
Finally, the results obtained using the TM construction scheme are in
excellent agreement with the equivalent results obtained by Trail and
Needs \cite{trail_needs1}. Any small differences could be attributed
to the different grids used, or the slight differences in the
self-consistent HF  procedure.

In Table~\ref{tab:tab_EA}, we show a similar study for the electron
affinity of the first- and second-row elements. The electron affinity
is the difference in energy between the neutral atom and the
negatively charged ion in their ground state configurations. The 
results agree to within $0.5$~mHa with previously published HF
electron affinities~\cite{EA}.

Again, these results show that the pseudopotentials have good
transferability properties.  Also, we note here that the HF
pseudopotentials generally bind an extra electron more strongly than
the DFT-based pseudopotentials. This is why the study of the electron
affinity is feasible.

We report the results of excitation energies in
Table~\ref{tab:tabexc}.  This is the difference in energy between the
ground state configuration shown in the first column and the two
excited state configurations shown in the third column. As before, the
pseudopotential results are in good agreement with the all-electron
values. The absolute average deviations from the all-electron values
are consistent across the different pseudopotential schemes, and is
less than 1.2~mHa. The largest deviation is with carbon and
is $\approx$5--6~mHa (the excitation energy between the
$^3P$ and $^5S$ states).

\begin{table}[t]
\caption{
\label{tab:tabrc2}
The parameters for the pseudopotentials used in the dimer calculations. The
cutoff radii  are in atomic units and $q_{c}^{\mathrm{max}}$ is in
$\mathrm{Ry}^{1/2}$.}
\begin{ruledtabular}
\begin{tabular}{lcccc}
\multicolumn{1}{r}{Atom} & $r_{cs}$ & $r_{cp}$ & $r_{cd}$ &  $q_{c}^{\mathrm{max}}$ \\ \hline
N       & 0.91 &  0.91 &  0.91 & 10.6 \\
P       & 1.58 & 1.58 & 1.90 & 5.5  \\
Cl      & 1.68 & 1.68 & 1.90 & 6.2  \\
\end{tabular}
\end{ruledtabular}
\end{table}   

\begin{figure}[t]
\includegraphics[width=3.4in,clip]{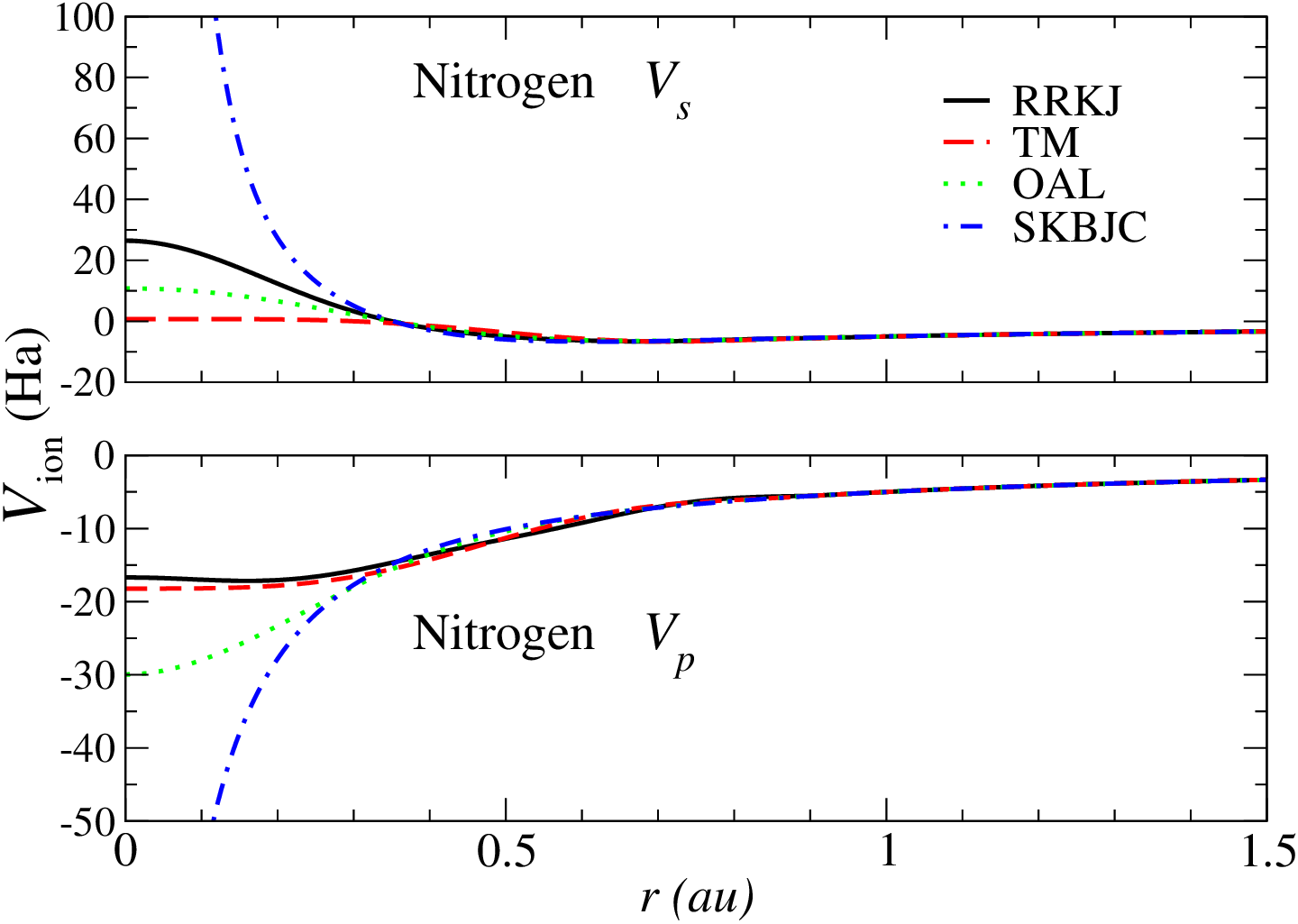}
\includegraphics[width=3.4in,clip]{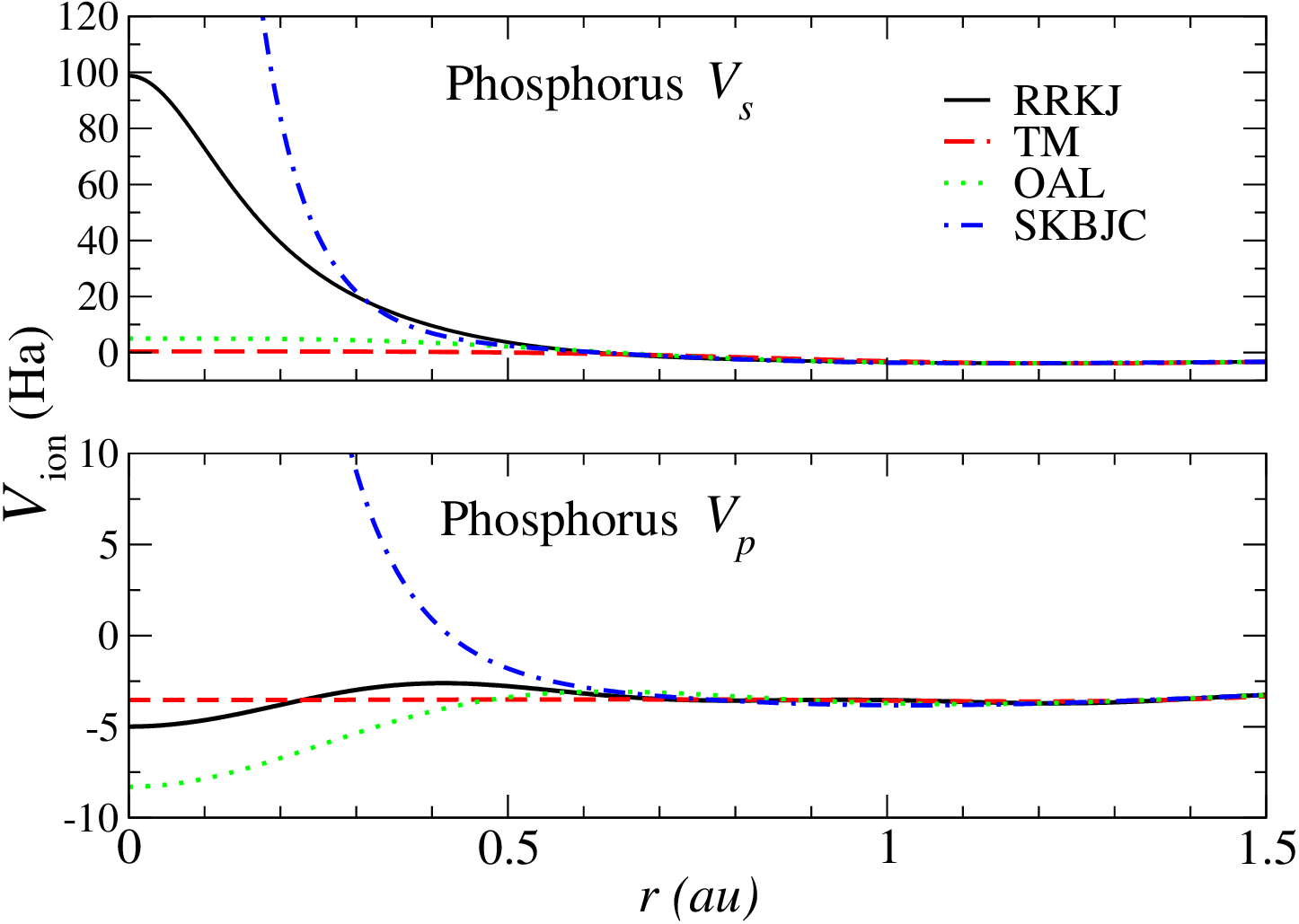}
\caption{\label{fig:Vi} (Color online) Comparison of several ionic
pseudopotentials for nitrogen and phosphorus. We show pseudopotentials
obtained using the RRKJ, TM, OAL \cite{lester_psp}, and the
SBKJC \cite{EMSL} methods. For the RRKJ and TM methods, the $r_c$
values are the same as those in Table~\ref{tab:tabrc2}.
We show only the $s$ and $p$ potentials. }
\end{figure}

\begin{figure}[t]
\includegraphics[width=3.2in,clip]{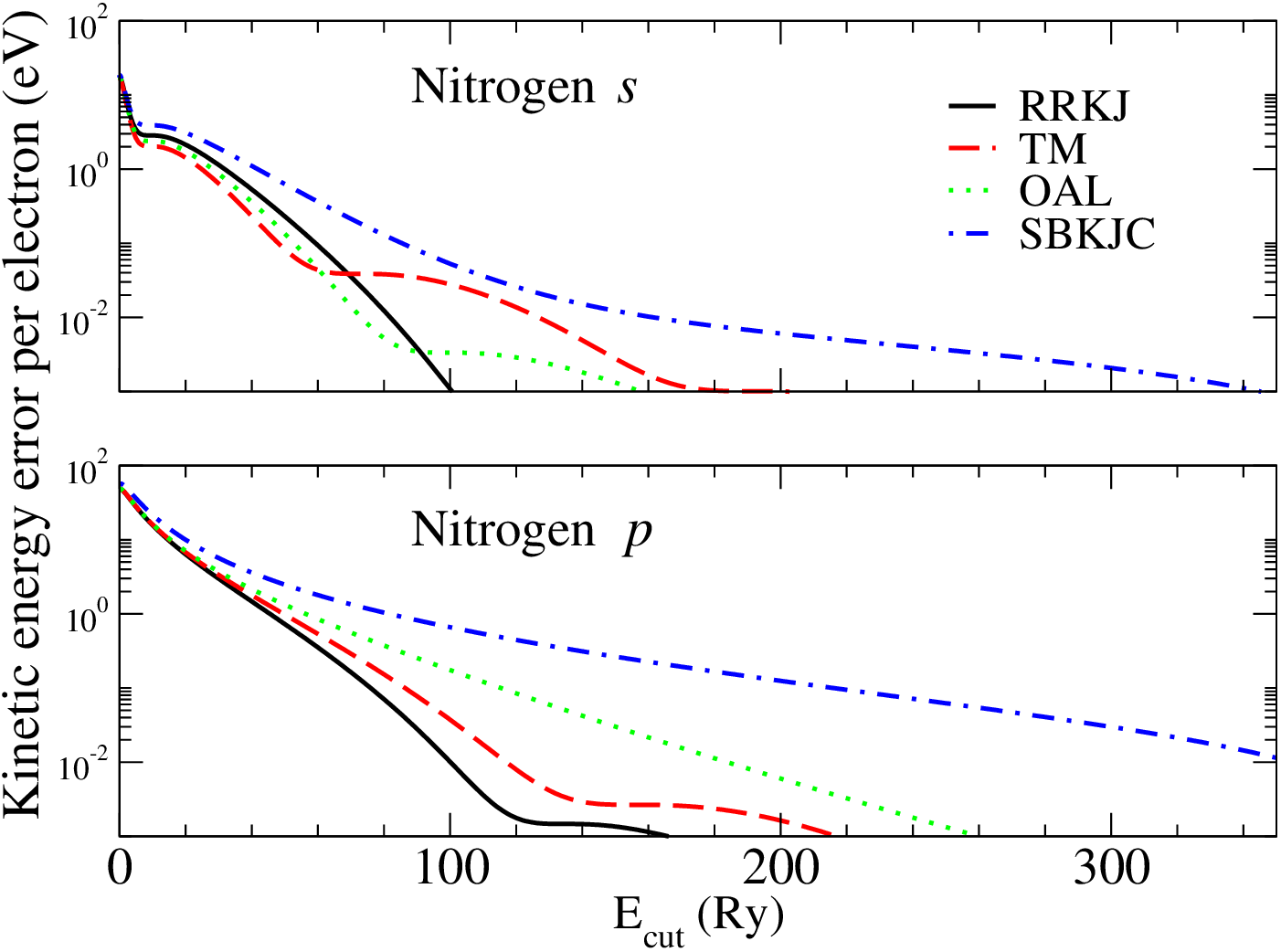}
\includegraphics[width=3.3in,clip]{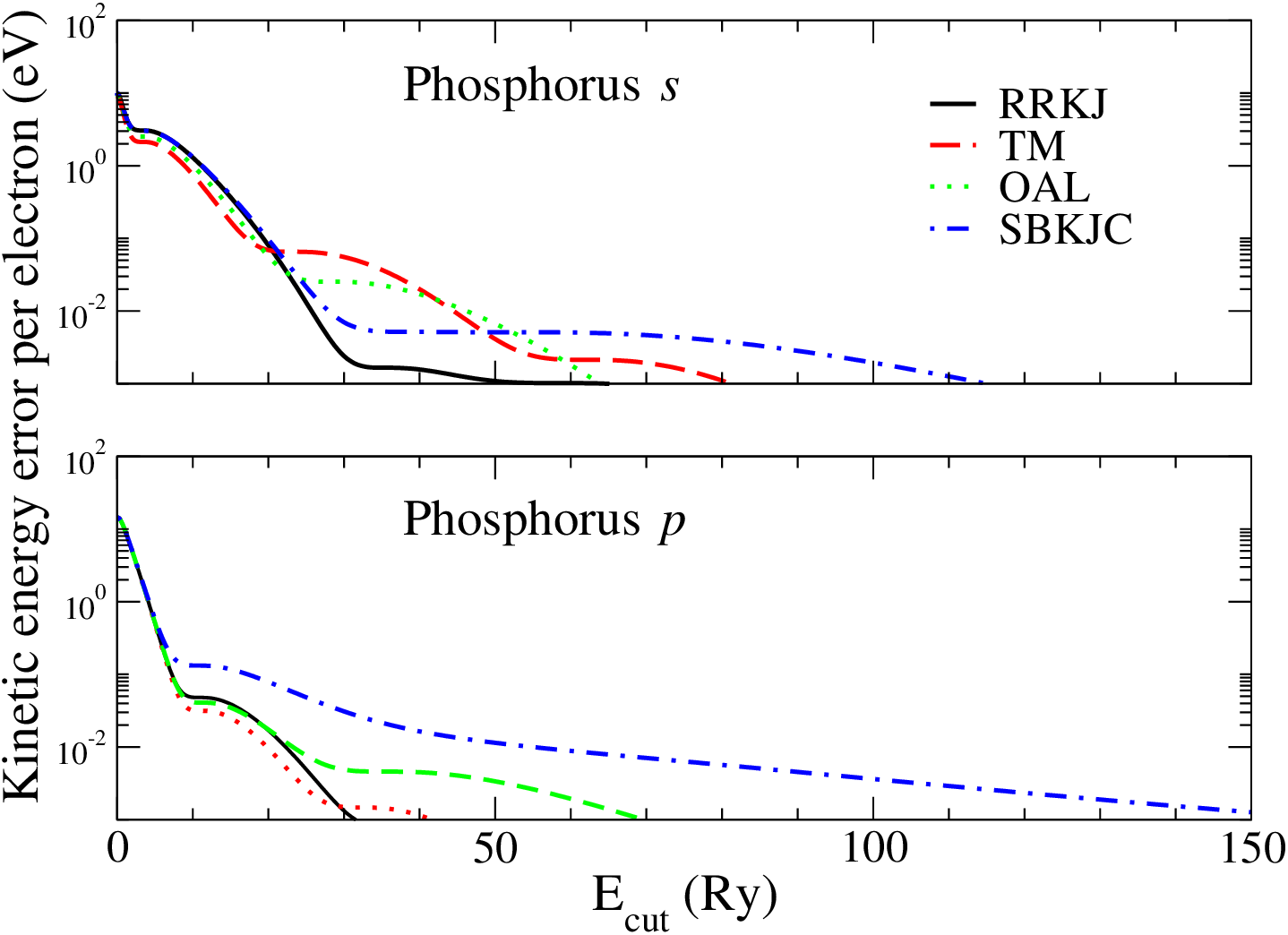}
\caption{\label{fig:KE} (Color online) Same as Fig.~\ref{fig:Vi},
  except that we show the residual kinetic energy convergence
  (logarithmic scale) for the potentials calculated using
  Eq.~(\ref{eq:resid_ke}). }
\end{figure}

\begin{table}[t]
\caption{Planewave basis cutoff (in Ry) for nitrogen and
  phosphorus pseudopotentials for several pseudopotentials
  methods. The planewave basis cutoff is estimated using
  Eq.~(\ref{eq:resid_ke}) for a residual kinetic energy of $\Delta
  T_{nl} \approx 5$~meV/electron, which can also be read from
  Fig.~\ref{fig:KE}.}
\label{tab:tabecps}
\begin{ruledtabular}
\begin{tabular}{lcc}
\multicolumn{1}{l}{} & \multicolumn{2}{c}{E$_{cut}$(Ry)}\\ 
 & N &  P \\ \hline
RRKJ           & 112  & 30 \\
TM             & 145  & 50 \\
OAL            & 210  & 55  \\
SBKJC          & 375  & 90  \\
\end{tabular}
\end{ruledtabular}
\end{table}

\begin{table}[bth]
\caption{
\label{tab:tab_molecule} Dissociation
energy $D_e$ in eV, bond length $R_e$ in Bohr, and harmonic frequency
$\omega_e$ in cm$^{-1}$ for several dimers as calculated using density
functional LDA and HF theories. We compare both the
all-electron and the pseudopotential results. For each method, we
generated the RRKJ pseudopotentials using the same level theory
i.e. LDA and HF, respectively.}
\begin{ruledtabular}
\begin{tabular}{llllcll} 
        &               &\multicolumn{2}{c}{LDA}                     & & \multicolumn{2}{c}{HF} \\
        &               &\multicolumn{1}{c}{AE} & \multicolumn{1}{c}{PSP} & & \multicolumn{1}{c}{AE} & \multicolumn{1}{c}{PSP}\\
\hline
N$_2$   &       $D_e$   &11.61 &10.96 &    & 5.10 & 4.92  \\        
        &       $R_e$   & 2.07 &2.06  &    & 2.01 & 2.02 \\
        &       $\omega_e$&2388&2380   &    & 2720  & 2722   \\ 
\\
P$_2$  &       $D_e$           &6.25  &6.02        &   &1.70    &1.70  \\  
       &       $R_e$           &3.57  &3.57        &   &3.50    &3.50  \\
        &       $\omega_e$      & 794   &788        &   & 912    & 905  \\      
\\
Cl$_2$  &       $D_e$          &3.62   &3.55   &   & 0.84& 0.84 \\
        &       $R_e$          &3.74   &3.74    &   & 3.73   & 3.72 \\
        &       $\omega_e$     &561     &559        &   & 610     & 613       \\        
\end{tabular}
\end{ruledtabular}
\end{table}

\section{Study of several dimers}

In this section, we apply our HF pseudopotentials in a study of three
dimers and compare them with all-electron calculations. The
pseudopotential calculations are done using a HF planewave basis code
\cite{cpmd}, and the all-electron results are obtained using a
Gaussian basis code \cite{gaussian}. For comparison, we have also
generated LDA \cite{lda} pseudopotentials using exactly the same
parameters as those of the HF pseudopotential, and we compare them
with their all-electron counterparts on an equal footing with the HF
results. In our all-electron and pseudopotential results we used the
Perdew-Wang \cite{lda} flavor of LDA. We did not include a non-linear
core correction (NLCC)\cite{nl_core,nl_core_porezag} in our LDA
pseudopotentials for LDA/HF comparison purposes. The pseudopotential
DFT calculations are carried out using ABINIT \cite{Abinit}. The
all-electron calculations are performed using the
correlation-consistent cc-pV5Z basis set for N and cc-pV(5+d)Z for P
and Cl \cite{dunning}. The differences between the quadrupole and
quintuple basis sets are negligible; e.g., in the binding energies the
differences are less than 0.02~eV. We also verified these values using
\mbox{6-311++G(3df,3pd)} basis sets.

In Table~\ref{tab:tabrc2}, we show the cutoff radii we used to
generate the RRKJ pseudopotentials for this study. These values, which
are different from those of Table~\ref{tab:tabrc}, are chosen to
accommodate the bond lengths of these dimers without core overlap when
possible.  In the case of P and Cl, the $d$ core cutoff radius is
allowed to extend beyond half of the dimer bond length to make softer
pseudopotentials. The atomic reference configurations are the same as
before.

Before presenting the molecular results, we comment on the planewave
convergence of the RRKJ pseudopotentials in comparison to other
pseudopotentials or ECPs.  We examined the pseudopotentials of
nitrogen and phosphorus as representatives of the first- and
second-row elements.  Figure~\ref{fig:Vi} shows the RRKJ, TM, OAL
\cite{lester_psp}, and SBKJC \cite{EMSL} pseudopotentials in real
space for the two elements. We did not include a $d$-channel in these
plots because neither the SBKJC nitrogen pseudopotential nor the OAL
potentials have a $d$-channel. For the RRKJ and TM pseudopotentials,
we used the cutoff radii as shown in Table~\ref{tab:tabrc2}.  Note
that the SBKJC ECP has a divergence at the origin, and is thus not
suitable for planewave calculations. More generally, direct
examination of pseudopotentials in real space gives little information
about the planewave basis needed for converged eigenstates and
eigenvalues.

One way to monitor the size of the planewave basis needed is to look
at the residual kinetic energy convergence \cite{RRKJ} of the
pseudowavefunctions, which we show in Fig.~\ref{fig:KE}. The residual
kinetic energy is calculated as shown in Eq.~(\ref{eq:resid_ke}) by
varying $q_c$ (we plot it against $q_{c}^{2}$ which is the cutoff
energy, for convenience). The planewave basis cutoff for each
pseudopotential is determined by the potential channel requiring the
largest cutoff energy.  The estimated planewave basis cutoffs based on
Fig.~\ref{fig:KE} are summarized in Table~\ref{tab:tabecps}.  As can
be seen, RRKJ pseudopotentials give the smallest planewave basis
cutoffs.

We summarize the results for the spectroscopic properties of the
dimers in Table~\ref{tab:tab_molecule}. The LDA and HF results for the
equilibrium bond length and the harmonic vibrational frequencies of
the dimers are shown to be well reproduced by both types of
pseudopotentials. In the LDA dissociation energies there is a large
discrepancy between all-electron and pseudopotential results for
N$_2$, and to a lesser extent in P$_2$ and even much less in Cl$_2$.
The HF pseudopotential dissociation energy of N$_2$ shows a
significant deviation from the all-electron HF result, whereas for
P$_2$ and Cl$_2$, the results are the same to within $0.01$~eV.  A
similar difference between all-electron and pseudopotential results
was also seen in Ref.~\cite{engel_exact_xc} using the OPM method.

Most of errors in the LDA results are due to linear descreening of the
pseudopotential, which is typically repaired by including a NLCC in
the target calculation. Furthermore, because all three dimers are
closed shell, the linear descreening effects are largest in the
isolated atom calculations and therefore mostly affect the
dissociation energies. As shown by Porezag \etal
\cite{nl_core_porezag}, these errors are larger for first row species
compared to heavier atoms and increase with more unpaired electrons,
explaining why N$_2$ has the largest error (first row, and 3 unpaired
spins) and why Cl$_2$ (second row, 1 unpaired spin) has the smallest.
Our LDA all-electron and pseudopotential energies are in excellent
agreement with previous results \cite{engel_exact_xc} without a NLCC.
This study also showed that the all-electron results are in good
agreement with the pseudopotentials values after including a NLCC,
indicating that this is the dominant error for LDA pseudopotentials.

At the HF level there is also an error in the exchange contribution to
the pseudopotential since the terms arising from core-valence
interactions in the atomic reference state are frozen into the core
and only valence-valence exchange terms are subtracted during the
descreening step. As shown by our results and others \cite{shirley,
hock-engel}, this type of descreening error is much smaller in
magnitude in HF pseudopotentials compared to LDA. This would suggest
that HF derived pseudopotentials are advantageous over those from LDA,
at least in better descreening of the core-valence contributions.

\section{Summary and conclusion}

Constructing HF pseudopotentials suitable for planewave calculations
is a non-trivial task considering the extreme nonlocality of the
exchange potential. Pseudopotentials constructed using a typical
norm-conserving procedure develop a non-local and a non-Coulombic tail
for large $r$. This would generally lead to erroneous results,
especially in solid calculations. Several schemes have been introduced
to cure the long tail behavior which lead to a small violation of the
conservation of the core charge
\cite{bk_exact_xc,engel_exact_xc,trail_needs1}.

In this study, we present soft-core HF pseudopotentials constructed
using the RRKJ procedure which optimizes the potentials to yield rapid
planewave basis cutoff convergence. The long tail behavior is fixed
using a self-consistent procedure following that of Trail and Needs
\cite{trail_needs1}, which leads to a negligibly small violation of
norm-conservation.  These pseudopotentials are applied in HF
calculations of several atomic properties yielding results in good
agreement with the all-electron values.  We also apply them in the
study of the dissociation energies, equilibrium bond lengths, and
frequency of vibrations of several dimers, using a HF planewave code
\cite{cpmd}. The all-electron and pseudopotential results are in
agreement with each other, and the values are consistent with a
similar comparison done using LDA pseudopotential and LDA all-electron
calculations.

Generation of HF based pseudopotentials has been released in version
3.0 of the GPL package OPIUM \cite{opium} and is available for
download.

\section{Acknowledgments:}

This work is supported by ONR grants N000140110365 and N000140510055,
by the Department of Energy Office of Basic Energy Sciences grants
DE-FG02-07ER15920 and DE-FG02-07ER46365, by the Air Force Office of
Scientific Research, Air Force Materiel Command, USAF, Grant
FA9550-07-1-0397 and by the NSF grant EAR-0530813.  We are grateful to
H. Krakauer and S. Zhang for many valuable discussions.


\begin{thebibliography}{10}
\bibitem{logdernote1} As in Ref.~\onlinecite{trail_needs1}, the
deviation in the logarithmic derivative is computed neglecting
changes in the Hartree and exchange potentials.
\bibitem{logdernote2} The logarithmic derivative is the ratio between
the  wavefunction and its slope, thus comparisons between
logarithmic derivatives can be misleading if either value is close to
zero. In all cases shown in Fig.~4, the magnitude of the wavefunctions
and slopes ranged between $0.02-2.2~ {\rm{Bohr}}^{-3}$ and
$0.01-1.6~ \rm{Bohr}^{-4}$, respectively.
\end{thebibliography}
\end{document}